**Review of a Heaviside step sequence function and the recursive Heaviside step sequence function for modeling human mental state..**

Changsoo Shin[1]

Emeritus Professor, Department of Energy resources Engineering, College of Engineering, Seoul National University, email: cssmodel@snu.ac.kr1. co-first author and corresponding author

## Abstract

In this paper, we define a novel recursive Heaviside step sequence function and demonstrate its applicability to modeling human mental states such as thought processes, memory recall, and forgetfulness. By extending the traditional Heaviside step function, which typically represents binary transitions, into a recursive sequence framework, we introduce a dynamic model that better captures the complexities of cognitive states. Furthermore, the recursive Heaviside step sequence function approximates solutions to a multidimensional inviscid advection equation, offering a unique mathematical perspective on the evolution of mental states over time. This continuous model, combined with the recursive delta sequence function, provides a comprehensive approach to exploring how memories and thoughts emerge, evolve, and fade. Through this approach, we propose that mental states can be expressed as time series functions, and the selection of the parameter N reflects individual variability in mental processing, influenced by external environments and internal experiences. We also discuss the implications of this framework for understanding human cognition and potential limitations due to modern technological constraints in replicating such processes.

## Keywords:

Recursive Heaviside step sequence function, delta sequence function, multidimensional advection equation, mental state potential, cognitive modeling.

## Introduction

This paper explores the application of recursive Heaviside step sequences to model human mental states, providing a novel mathematical approach to understanding complex cognitive phenomena. The recursive Heaviside step function builds upon the traditional Heaviside step function, which is widely used to represent binary transitions—such as 'on/off' or 'conscious/unconscious' states. However, human cognition is far from binary, involving layers of thought processes, memory recall, and various degrees of awareness and forgetfulness.

To address this complexity, we introduce a recursive structure to the Heaviside function, transforming it into a sequence that models the gradual and interconnected nature of cognitive events. By incorporating recursive delta sequences, we further refine this model, allowing for the representation of rapid transitions between mental states, akin to the delta function's role in capturing instantaneous change.

The recursive Heaviside step sequence function provides a flexible framework for describing mental states as they evolve over time. The function’s recursive nature allows for the modeling of both conscious thought and the underlying processes that influence it, such as subconscious associations and forgotten memories. This approach also offers a way to mathematically differentiate between varying levels of cognitive engagement, where the parameter N reflects the complexity of mental processing—ranging from simple, isolated thoughts to deeply interwoven memories and experiences.

One of the key contributions of this paper is the application of this recursive function to a multidimensional inviscid advection equation. This equation typically models the transport of quantities like fluid dynamics or energy across space, but here, it is applied to mental states, showing how thoughts and memories 'move' and evolve over time. When N approaches infinity, the recursive Heaviside step function fully satisfies the multidimensional advection equation, while smaller values of N reflect approximate or incomplete mental states.

By extending these mathematical tools to the realm of cognitive science, we propose that mental states can be viewed as time series functions, where different moments of experience and thought are interconnected through a recursive process. This perspective not only deepens our understanding of human cognition but also presents opportunities for future research in both the fields of applied mathematics and cognitive science.

In order to fully understand the recursive Heaviside step sequence function and its applications, we must first define the key mathematical components used throughout this paper: the Heaviside sequence function and the delta sequence function. These sequences serve as the foundation for modeling human cognitive states in a dynamic, continuous framework.

### Heaviside Sequence Function

The classical Heaviside step function H(t) is defined as a unit step function where H(t)=1for t≥ and H(t)=0, which, while useful in physical systems, is too simplistic for representing the nuances of human cognition. To address this limitation, we extend the Heaviside step function into a **Heaviside sequence function**, defined as:

$$H_N(t) = \lim_{k \to N} \frac{1}{1+e^{-2kt}} \qquad (1-1)$$

This formulation smooths the transition between states, creating a gradual shift from 0 to 1, which is more representative of mental states that are rarely purely binary. Here, N serves as a parameter controlling the rate of transition, and as N→∞, the function approaches the classical Heaviside step function. Figure (1) shows four kinds of Heaviside sequences where the green line corresponds to N=1.0,the blue line to N=0.5, the yellow line to N=0.25,and the red to N=0.1.

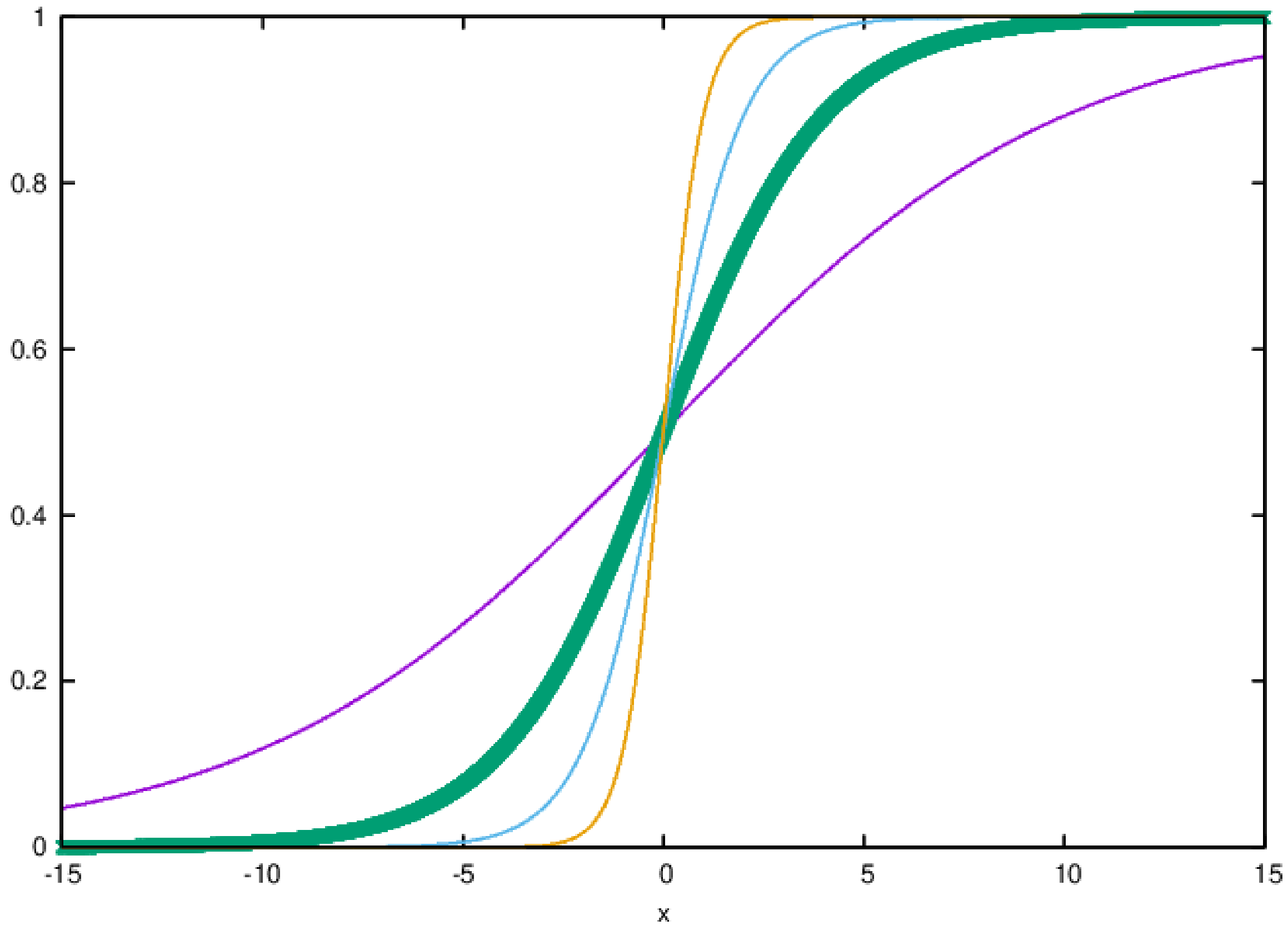

Fig.(1). Heaviside sequences: the orange line corresponds to N=1, the blue line to N=0.5,the green line to N=0.25,and the Purple to N=0.1.

Similarly, the delta function δ(t)), which represents an instantaneous change, is adapted into a **delta sequence function** for more nuanced representation of cognitive shifts. The delta sequence function is expressed as:

$$\delta_N(t) = \lim_{k \to N} \frac{2ne^{-2nt}}{(1 + e^{-2nt})^2} \qquad (1-2)$$

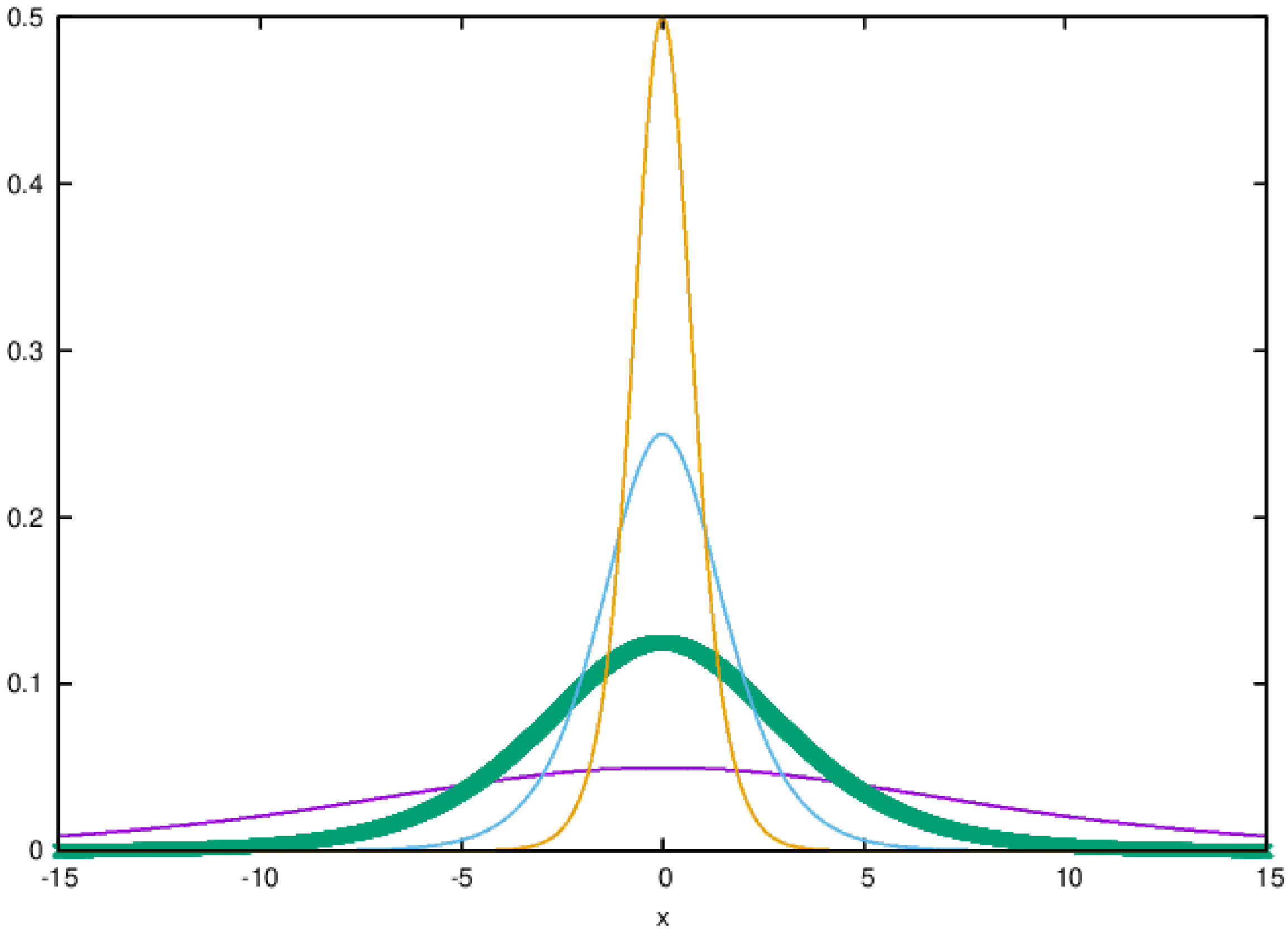

Fig.(2). Delta sequences: the orange line corresponds to N=1, the blue line to N=0.5,the green line to N=0.25,and the purple to N=0.1.

This sequence captures sharp transitions in mental states, such as sudden realizations or rapid shifts in focus. Like the Heaviside sequence function, the delta sequence function approaches the classical delta function as N→∞. The recursive nature of these functions allows for modeling not only individual mental events but also the interactions between multiple cognitive processes.

Figure (2) shows four kinds of delta sequences where the green line corresponds to N=1.0,the blue line to N=0.5, the yellow line to N=0.25,and the red to N=0.1

**Recursive Heaviside Step Sequence Function**

The recursive Heaviside step sequence function is constructed by embedding multiple Heaviside sequences within each other, effectively capturing layers of thought, memory, and cognitive activity. For example, the function can be written as:

$$u(t) = 1 - H_{0.1}\left(-t + \tau_1 H_1\left(-t + \tau_2 H_2\left(-t + \tau_3 H_5(-t + \tau_4)\right)\right)\right) \qquad (3-1)$$

where $\tau_1 \leq \tau_2 \leq \tau_3 \leq \tau_4$. If $\tau_1 = \tau_2 = \tau_3 = \tau_4$, equation(3) becomes

$$u(t) = 1 - H_{0.1}\left(-t + \tau_1 H_1\left(-t + \tau_1 H_2\left(-t + \tau_1 H_5(-t + \tau_1)\right)\right)\right) \qquad (3-2)$$

Furthermore, If N approaches to $\infty$, equation(3-2) becomes

$$u(t) = 1 - H_{\infty}\left(-t + \tau_1 H_{\infty}\left(-t + \tau_1 H_{\infty}\left(-t + \tau_1 H_{\infty}(-t + \tau_1)\right)\right)\right) \qquad (3-3)$$

If this nested loop is embedded infinitely, equation(3-3) becomes

$$u(t) = 1 - H_{\infty}\left(-t + \tau_1 H_{\infty}(-t + \tau_1 H_{\infty} ---------------)\right) \qquad (3-4)$$

Equation(3-4) is related with "creation of universe"(Shin,2016 and Sin and Kim(2019)), In Shin(2016)'s paper, he defined u(t) as a divine God's potential. Here after, we call u(t) a mental potential.

This recursive structure models how thoughts build upon one another over time. The parameter N in each sequence reflects the strength or intensity of each cognitive event, with larger values representing more pronounced or impactful mental states. When N is large, the cognitive state is clearly defined, but as N decreases, the state becomes less distinct, modeling processes such as forgetting or uncertainty.

---

## Applications to Human Cognition

The recursive Heaviside step sequence function and the delta sequence function offer powerful tools for modeling various aspects of human cognition. In this section, we explore how these functions can represent key cognitive processes: **thought formation**, **memory recall**, **forgetfulness**, and **the association of ideas**.

### Thought Formation

In human cognition, thoughts rarely emerge instantaneously. Rather, they develop gradually, shaped by prior experiences and external stimuli. The recursive Heaviside step sequence function effectively captures this process, where each embedded sequence reflects a layer of cognitive buildup. For instance, in the example function:

$$u(t) = 1 - H_{0.1}\left(-t + H_1\left(-t + 2H_2\left(-t + 3H_5(-t + 4)\right)\right)\right) \qquad (4)$$

Each nested H(t) represents a previous mental event influencing the current state. This recursive nature models how thoughts are formed over time, incorporating past memories (represented by $\tau$ and current inputs).

### Memory Recall

Memory recall can be interpreted as the “activation” of previously stored mental states. In the recursive Heaviside framework, this process is modeled by the derivative of the Heaviside sequence function with respect to time or event time($\tau_i$) When recalling a memory, the derivative reflects the influence of past experiences on the present mental state: Let’s consider a simple mental potential given as below

## Memory Recall in the Recursive Heaviside Framework

Memory recall can be interpreted as the "activation" of previously stored mental states. In the recursive Heaviside framework, this process is modeled by the derivative of the Heaviside sequence function with respect to time or event time ($\tau_i$)When recalling a memory, the derivative reflects the influence of past experiences on the present mental state.

Let’s consider a simple mental potential, represented as:

$$\mathrm{u}_4 = 1 - H_N(-t+4) \qquad (5-1)$$

By taking the derivative of equation (5-1) and defining this as the state of remembering, we obtain:

$$\left|\frac{\partial \mathrm{u}_4}{\partial \mathrm{t}}\right| = \delta_{\mathrm{N}}(-t+4) \qquad (5-2)$$

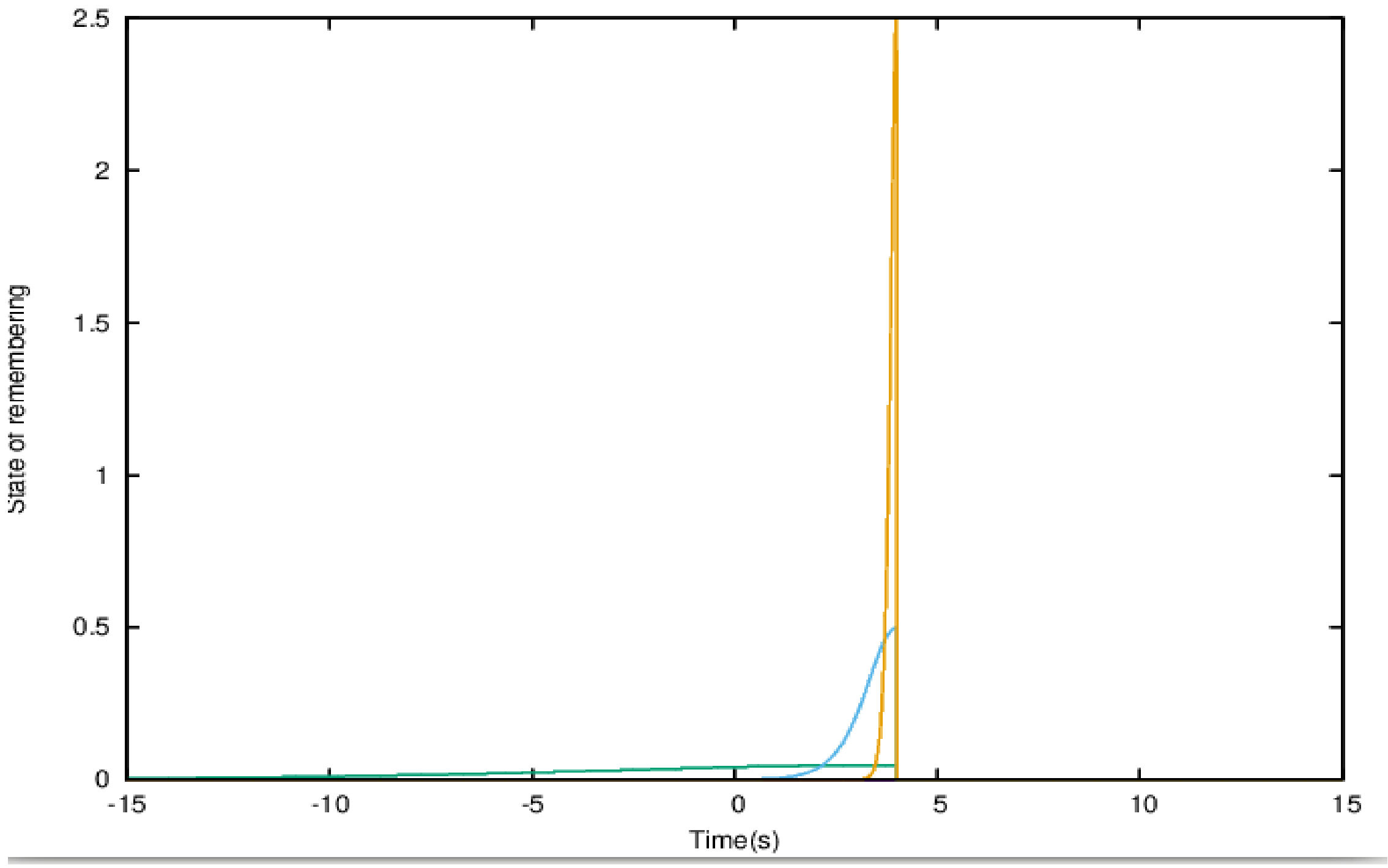

Fig.(3). State of Remembering: orange correspond to N=15, the blue to N=2, the green to N=1

,and purple to N=0.1.

Here, we note that t must be less than or equal to 4, in accordance with the definition of the Heaviside step function. This result, expressed as a half-side Delta sequence, captures the sharp transition that occurs when a memory is recalled. This reflects the fact that memories are always recalled in the present—"we never remember in the past"—as each recollection reactivates previous layers of thought, contributing to the current cognitive state.

The recursive structure of this framework allows for modeling not only the recall of a single event but also how multiple memories can overlap or intertwine during the recall process. This approach adds a new dimension to understanding memory by quantifying the dynamics of recall and fading over time, paralleling established cognitive models of memory decay, such as Ebbinghaus's forgetting curve (Ebbinghaus, 1885) and the Atkinson-Shiffrin model (Atkinson & Shiffrin, 1968).We note that t should be less than 4 or equal 4 because of definition of Heaviside step function.

We display equation(5-2) in Figure(3) as a function of N. This half side delta sequence (we never remember in the past)captures the sharp transition that occurs when a memory is recalled, where previous layers of thought are reactivated, contributing to the present cognitive state. The recursive structure allows for modeling not only the recall of a single event but also how multiple memories can overlap or intertwine during the recall process. This approach parallels established cognitive models of memory decay, such as Ebbinghaus's forgetting curve (Ebbinghaus, 1885) and the Atkinson-Shiffrin model (Atkinson & Shiffrin, 1968), but introduces a new dimension by explicitly quantifying memory recall and fading over time."

### Forgetfulness

The process of forgetting can be modeled by allowing the parameter N in the recursive Heaviside step sequence function to decrease over time. As N decreases, the transitions between states become less sharp, reflecting the gradual fading of memories. In the extreme case where N is small, the recursive function no longer satisfies the multidimensional advection equation, representing a state where the memory has decayed to the point of irrelevance.

$$u(t) = 1 - H_{0.1}\left(-t + H_1\left(-t + 2H_2\left(-t + 3H_5(-t + 4)\right)\right)\right) \tag{6}$$

As N diminishes, the contribution of each past event becomes weaker, eventually leading to complete forgetting.

### Association of Ideas

Another key aspect of human cognition is the ability to associate seemingly unrelated ideas. In the recursive Heaviside step sequence model, associations are represented by the interactions between different nested sequences. For example, when recalling a memory associated with multiple events, the recursive structure allows for the reactivation of related memories, even if they were not consciously recalled at first. This is reflected in the layered nature of the sequence:

$$u(t) = 1 - H_{0.1}\left(-t + \tau_1 H_1\left(-t + \tau_2 H_2\left(-t + \tau_3 H_5(-t + \tau_4)\right)\right)\right) \quad (7)$$

The associations between ideas are captured by the recursive connections between different events $\tau_1$, $\tau_2$, $\tau_3$, and $\tau_4$ reflecting the interconnected nature of cognitive processes.

---

## Mathematical Modeling

In this section, we present the mathematical foundation for the recursive Heaviside step sequence function and its connection to multidimensional advection equations. This framework allows us to model the evolution of mental states as dynamic processes influenced by both time and external events.

### Recursive Heaviside Step Sequence and the Advection Equation

The recursive Heaviside step sequence function can be represented as a time series, capturing the development of cognitive states over time. The recursive nature of the function, embedded with layers of Heaviside sequences, allows it to approximate solutions to the n-**dimensional inviscid advection equation**. This equation is commonly used in fluid dynamics to describe the transport of a quantity, but here, we reinterpret it in the context of mental states.

The general form of the n-dimensional advection equation is written as:

$$\left[H_\infty(-t+\tau_1)\right]\left[(n-1)\left|\frac{\partial u_1}{\partial t}\right| + \frac{\partial u_1}{\partial \tau_1} + \frac{\partial u_1}{\partial \tau_2} + \frac{\partial u_1}{\partial \tau_3} + + + + \cdots + \frac{\partial u_1}{\partial \tau_n}\right] = 0 \quad (8-1)$$

As we progress through each subsequent layer of the recursive Heaviside function, the complexity of the advection equation increases. For example, the second and third layers of the recursive function are governed by:

$$\left[H_\infty(-t+\tau_2)\right]\left[(n-2)\left|\frac{\partial u_2}{\partial t}\right| + \frac{\partial u_2}{\partial \tau_2} + \frac{\partial u_2}{\partial \tau_3} + \frac{\partial u_2}{\partial \tau_4} + + + + \cdots + \frac{\partial u_1}{\partial \tau_n}\right] = 0 \quad (8-2)$$

$$\left[H_\infty(-t+\tau_3)\right]\left[(n-3)\left|\frac{\partial u_3}{\partial t}\right| + \frac{\partial u_3}{\partial \tau_3} + \frac{\partial u_3}{\partial \tau_4} + \frac{\partial u_3}{\partial \tau_5} + + + + \cdots + \frac{\partial u_1}{\partial \tau_n}\right] = 0 \quad (8-3)$$

.

.

These recursive equations describe how different layers of thought (represented by$u_1, u_2, u_3$, ,...) evolve over time, influenced by past events ($\tau_1, \tau_2, \tau_3$,...).

By inspection, the solutions are as follows:

$$u_1 = 1 - H(-t + \tau_1 H(-t + \tau_2 H(-t + \tau_3 H(-t + \tau_4 \ldots.. \quad (9-1)$$

$$u_2 = 1 - H(-t + \tau_2 H(-t + \tau_3 H(-t + \tau_4 H(-t + \tau_5 \ldots.. \quad (9-2)$$

$$u_3 = 1 - H(-t + \tau_3 H(-t + \tau_4 H(-t + \tau_5 H(-t + \tau_6 \ldots.. \quad (9-3)$$

Here, H represents the Heaviside step function, t is time, and $\tau_n$is the moment when an input is received by one of the human sensory organs. Time flows from left to right, where H(t)=1 for t≥0 and H(t)=0 for t<0.

The n-dimensional advection equations (8-1), (8-2), and (8-3) are exact solutions when N approaches infinity. When N is finite, these equations can be considered as approximations.

The recursive Heaviside step function was first introduced by Shin et al. (2014) to describe human mental states. However, the dilemma in their work was that human mental states were expressed as binary values (0 or 1). In their framework, consciousness was represented as 1 (alive/conscious), while unconsciousness (anti-consciousness) was represented as 0, when using the Heaviside step function and delta function.

### The Role of N in Cognitive Modeling

A key insight from this mathematical model is the role of the parameter N, which determines the smoothness and complexity of the transitions between mental states. When N approaches infinity, the recursive Heaviside step function satisfies the N-dimensional advection equation exactly. This scenario corresponds to a mental state that is well-defined, stable, and clearly connected to previous experiences.

However, when N is finite or small, the recursive Heaviside step function only approximates the solution to the advection equation. In this case, the mental state becomes less defined, which can be interpreted as cognitive $\delta_N(t)$uncertainty, forgetfulness, or a lack of clarity in thought. The selection of N is context-dependent and reflects the individual's experience and environmental factors, such as emotional state or external stimuli.

The delta sequence function$\delta_N(t)$ further refines this model by capturing rapid transitions or sudden realizations within the mental process. The derivative of the recursive Heaviside step function with respect to time t or event time $\tau_i$reflects shifts in cognitive states, allowing us to model both gradual and instantaneous changes in thought processes.

$$\left|\frac{\partial \mathrm{u}_4}{\partial \mathrm{t}}\right| = H_\infty(-t+4)\ \delta_{\mathrm{N}}(-t+4)$$

This derivative captures the transition from one mental state to another, illustrating how the mind shifts focus or recalls specific events over time, depending upon N.

Even though we recall an event happened 1 second ago, our remembering can be sharp for big N, vague for medium N, fading for small N, and forgetting where N approaches to 0, as shown in Figure(2).We note that Delta sequences can represent may states of our remembering.

### Time Series Representation of Mental States

By embedding the recursive Heaviside step sequence within the context of a time series, we can model human cognition as an evolving system. The mental state at any given moment is influenced not only by current inputs but also by previous experiences and thoughts, layered recursively within the function. This time-series model is continuous, reflecting the fact that human cognition is rarely static, constantly shifting as new thoughts and experiences occur.

For example, let us consider the mental state of a person who has experienced four different events. We can represent this mental state using a recursive Heaviside step sequence function as follows:

$$\mathrm{u(t)} = 1 - \mathrm{H}_{0.1}\left(-\mathrm{t} + \mathrm{H}_1\left(-\mathrm{t} + 2\mathrm{H}_2\big(-\mathrm{t} + 3\mathrm{H}_5(-\mathrm{t}+4)\big)\right)\right) \tag{10}$$

This equation assumes the person encountered events at 4, 3, 2, and 1 seconds prior. We can separate equation (4) into four sequence functions, as done in the previous section*:*

$$\mathrm{u}_4(\mathrm{t}) = 1 - \mathrm{H}_5(-t+4) \tag{11}$$

$$\mathrm{u}_3(\mathrm{t}) = 1 - \mathrm{H}_2\big(-\mathrm{t} + 3\mathrm{H}_5(-\mathrm{t}+4)\big) \tag{12}$$

$$\mathrm{u}_2(\mathrm{t}) = 1 - \mathrm{H}_1\left(-\mathrm{t} + 2\mathrm{H}_2\big(-\mathrm{t} + 3\mathrm{H}_5(-\mathrm{t}+4)\big)\right) \tag{13}$$

$$\mathrm{u}_1(\mathrm{t}) = 1 - \mathrm{H}_{0.1}\left(-\mathrm{t} + \mathrm{H}_1\left(-\mathrm{t} + 2\mathrm{H}_2\big(-\mathrm{t} + 3\mathrm{H}_5(-\mathrm{t}+4)\big)\right)\right) \tag{14}$$

Here, we refer to each ,$u_1(t), u_2(t), u_3(t)$ and $u_4(t)$ in equations (11), (12), (13), and (14) as **mental potentials**.

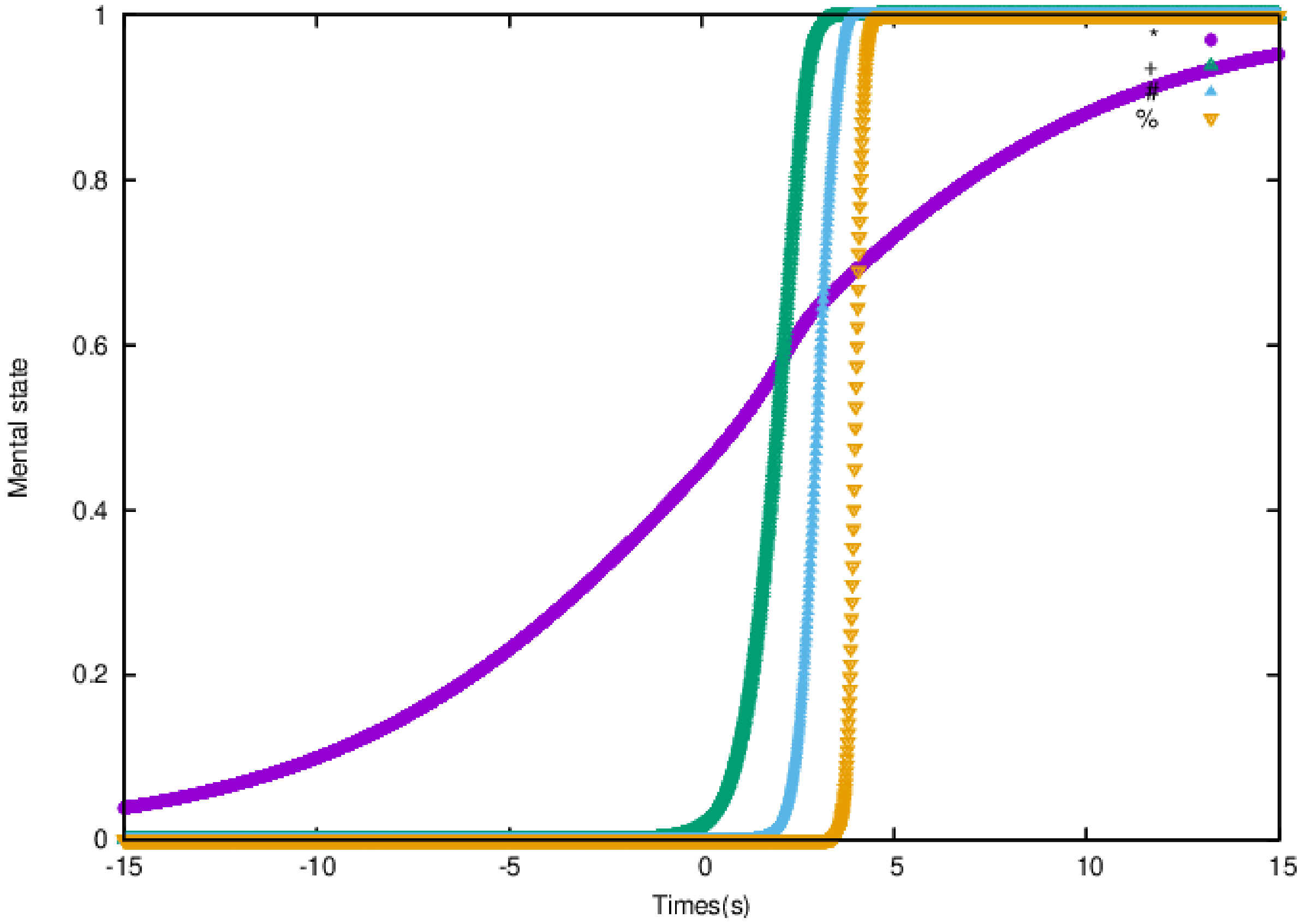

Fig.(4) Mental potentials: the orange line corresponds to equation (11), the blue line to quation(12),the green line to equation(13),and the purple to equation(14).

In Figure (4),the orange line corresponds equation(11),the blue line to equation(12),the Green line equation(13),and the purple line equation(14). All lines in Figure(3) represent the flow from left to right. This representation suggests that each event experienced by the five human sensory organs can be registered and stored as a function of the recursive Heaviside step sequence, as shown in Figure (3),depending on N. Here, we define the extent to which each individual internalizes an experience as free will. The value of N is assumed to vary according to external conditions, environmental factors, and other situations. Figure 5 shows the recursive Heaviside step function when N approaches infinity. We assume that the mental state of a human can be expressed by the recursive Heaviside step sequence (time series) function. However, under current technology, it is not feasible to construct such a storage device.

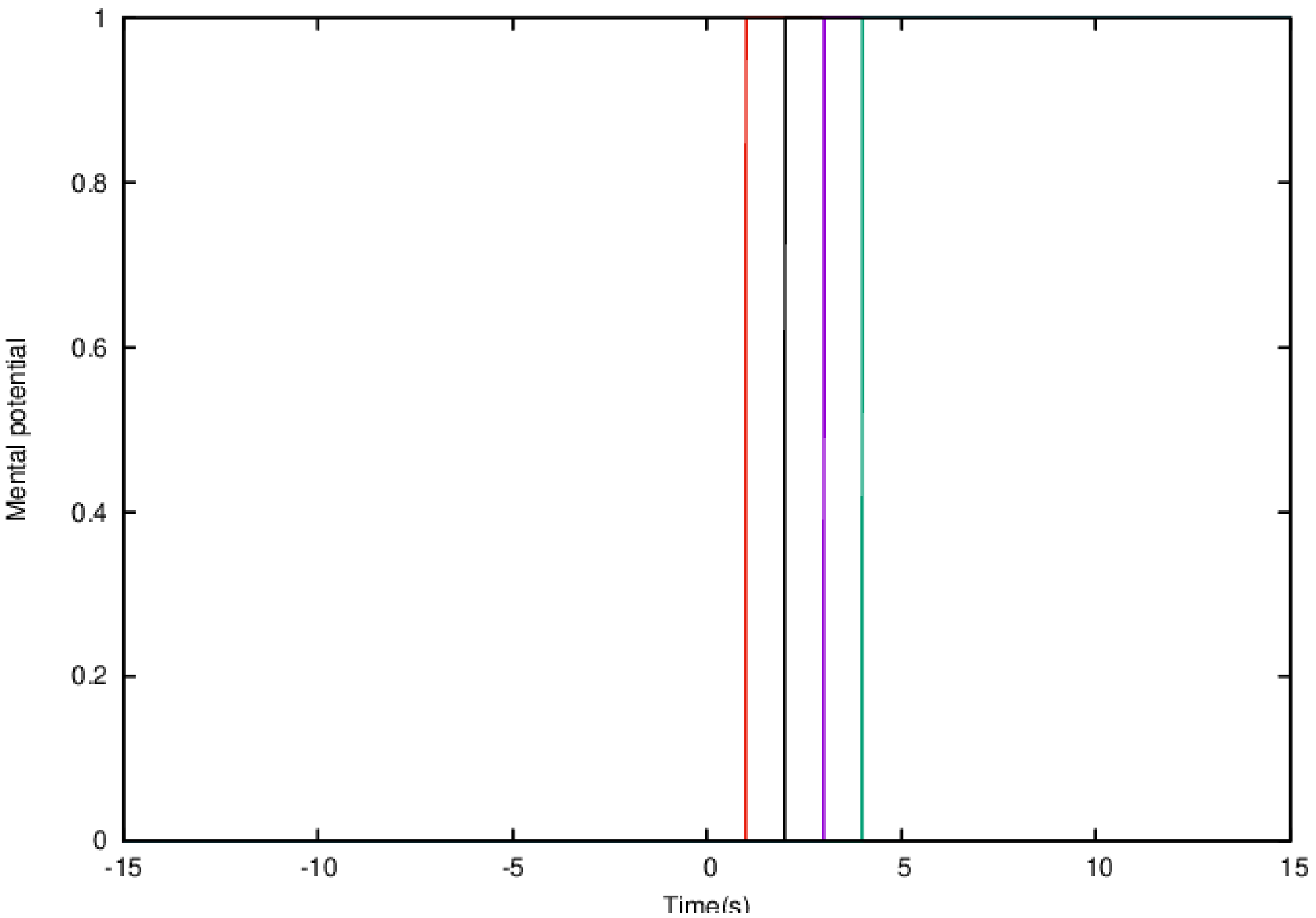

Fig.(5) Mental potentials: the orange line corresponds to equation (11), the blue line to equation(12),the green line to equation(13),and the purple to equation(14) when N approaches to infinity.

### Derivative of the Heaviside step sequence function with respect to time(t)

We can consider two types of derivatives: with respect to time (t) and with respect to each event time$\tau_i$ , where i=1,2,…,n. Taking the derivative of equations (11) to (14) with respect to time (t), we obtain the following:

$$\left|\frac{\partial u_4}{\partial t}\right| = H_\infty(-t+4)\ \delta_5(-t+4) \quad (15)$$

$$\left|\frac{\partial u_3}{\partial t}\right| = H_\infty(-t+3)\delta_2\left(-t+3H_5(-t+4)\right)\left(1+3\delta_5\left((-t+4)\right)\right) \quad (16)$$

$$\left|\frac{\partial u_2}{\partial t}\right| = H_\infty(-t+2)\delta_1\left(-t+2H_2\left(-t+3H_5(-t+4)\right)\right)\left(1+2\delta_2\left(-t+3H_5(-t+4)\right)\right) \quad (17)$$

$$\left(1+3\delta_5(-t+4)\right)$$

$$\left|\frac{\partial u_1}{\partial t}\right| = H_\infty(-t+1)\delta_{0.1}\left(-t+H_1\left(-t+2H_2\left(-t+3H_5(-t+4)\right)\right)\right) \quad (18)$$

$$\left(1+\delta_1\left(-t+2H_2\left(-t+3H_5(-t+4)\right)\right)\right)$$

$$\left(1+2\delta_2\left(-t+3H_5(-t+4)\right)\right)$$

$$\left(1+3\delta_5(-t+4)\right)$$

In this paper, we define "a thinking or remembering state or associating " as the derivative of the human mental state (potential) with respect to time (t).

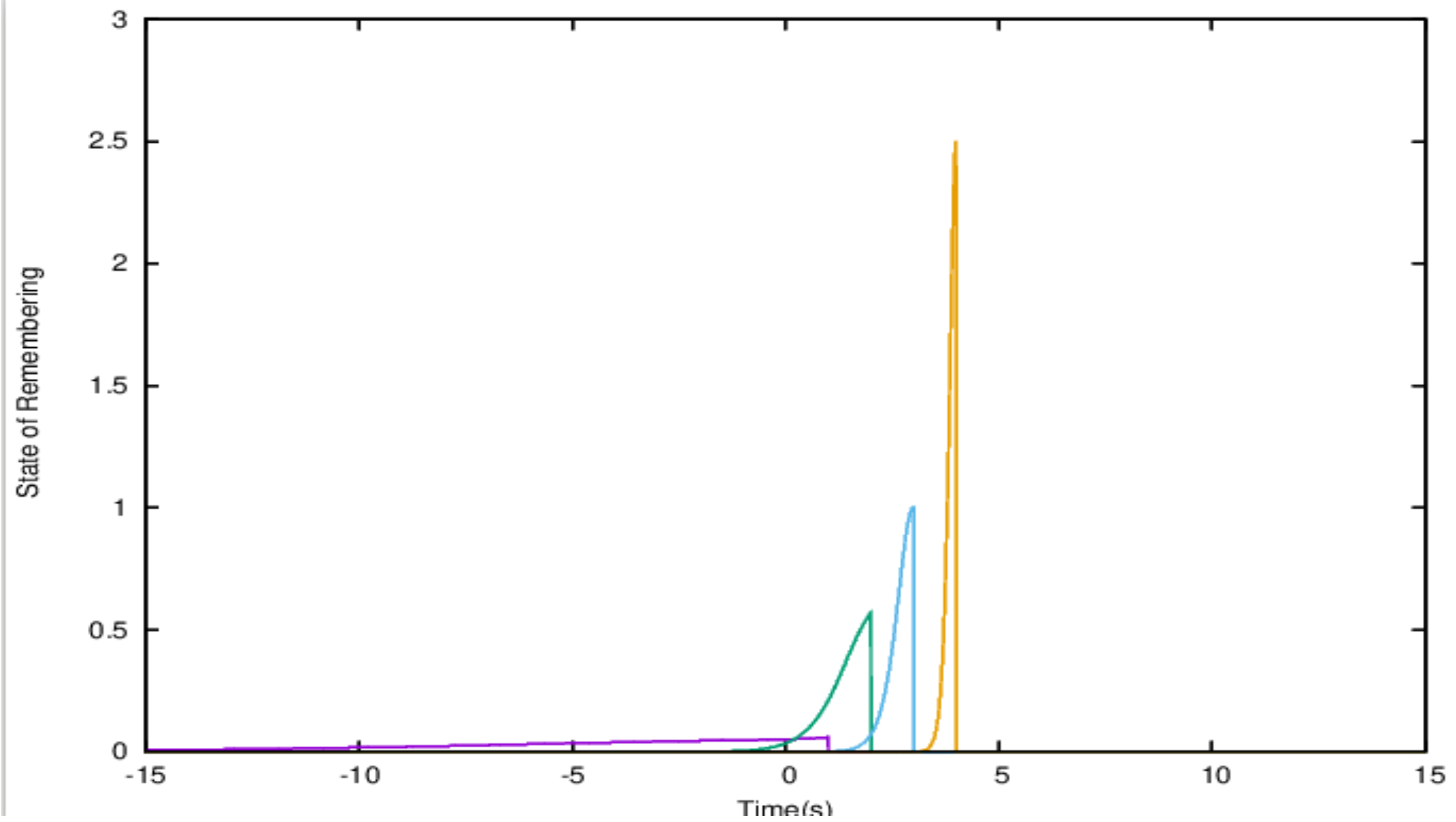

Fig.(6).State of remembering: the orange line corresponds to equation (15), the blue line to equation(16),the green line to equation(17),and the purple to equation(18).

In Figure 6, the orange line corresponds to equation (15), the blue line represents equation (16), the green line corresponds to equation (17), and the purple line shows a time series corresponding to equation (18). The orange line represents a simple response (thought), while the blue and green lines show a response in a broader state. On the other hand, the purple line represents a complex, changing state. This is because, as shown in equation (18), past experiences are intertwined and appear together.

As we decrease order of both the Heaviside sequence function and the Delta sequence function, the past events appeared in our mind. If we usually cannot remember order of Heaviside sequences exactly, our remembering may be close to truth or distorted. We may say that our brain is a derivative machine of the mental state potentials.

### Derivative of the Heaviside step sequence function with respect to the event time( $\tau_i$)

Equations (5), (6), (7), and (8) can be rewritten as a function of time and event time

$$u_4(t) = 1 - H_5(-t+\tau_4) \quad (19)$$

$$u_3(t) = 1 - H_2\left(-t+\tau_3 H_5(-t+\tau_4)\right) \quad (20)$$

$$u_2(t) = 1 - H_1\left(-t + \tau_2 H_2\left(-t + \tau_3 H_5(-t + \tau_4)\right)\right) \quad (21)$$

$$u_1(t)=1-H_{0.1}\left(-t+\tau_1 H_1\left(-t+\tau_2 H_2\left(-t+\tau_3 H_5(-t+\tau_4)\right)\right)\right) \quad (22)$$

When we take the derivative of equations (13), (14), (15), and (16) with respect to $\tau_i$ where i=1,2,3,4 we obtain

$$\frac{\partial u_4}{\partial \tau_4} = -H_\infty(-t + \tau_4)\delta_5(-t + \tau_4) \quad (23)$$

$$\frac{\partial u_3}{\partial \tau_3} = -H_\infty(-t + \tau_3)\delta_2\left(-t + \tau_3 H_5(-t + \tau_4)\right)H_5(-t + \tau_4) \quad (24)$$

$$\frac{\partial u_3}{\partial \tau_4} = -H_\infty(-t + \tau_3)\delta_2\left(-t + \tau_3 H_5(-t + \tau_4)\right)\tau_3\delta_5(-t + 4) \quad (25)$$

Similarly, we have for $u_2(t)$

$$\frac{\partial u_2}{\partial \tau_2} = -H_\infty(-t + \tau_2)\delta_1\left(-t + \tau_2 H_2\left(-t + \tau_3 H_5(-t + \tau_4)\right)\right)H_2\left(-t + \tau_3 H_5(-t + \tau_4)\right) \quad (26)$$

$$\frac{\partial u_2}{\partial \tau_3} = -H_\infty(-t + \tau_2)\delta_1\left(-t + \tau_2 H_2\left(-t + \tau_3 H_5(-t + \tau_4)\right)\right)H_5(-t + \tau_4) \quad (27)$$

$$\frac{\partial u_2}{\partial \tau_4} = -H_\infty(-t + \tau_2)\delta_1\left(-t + \tau_2 H_2\left(-t + \tau_3 H_5(-t + \tau_4)\right)\right)\tau_3\delta_5(-t + \tau_4) \quad (28)$$

Similarly, we obtain for $u_1(t)$

$$\frac{\partial u_1}{\partial \tau_1} = -H_\infty(-t + \tau_1)\delta_{0.1}\left(-t + \tau_1 H_1\left(-t + \tau_2 H_2\left(-t + \tau_3 H_5(-t + \tau_4)\right)\right)\right) \quad (29)$$

$$H_1\left(-t + \tau_2 H_2\left(-t + \tau_3 H_5(-t + \tau_4)\right)\right)$$

$$\frac{\partial u_1}{\partial \tau_2} = -H_\infty(-t + \tau_1)\delta_{0.1}\left(-t + \tau_1 H_1\left(-t + \tau_2 H_2\left(-t + \tau_3 H_5(-t + \tau_4)\right)\right)\right) \quad (30)$$

$$H_2\left(-t + \tau_3 H_5(-t + \tau_4)\right)$$

$$\frac{\partial u_1}{\partial \tau_3} = -H_\infty(-t + \tau_1)\delta_{0.1}\left(-t + \tau_1 H_6\left(-t + \tau_2 H_7\left(-t + \tau_3 H_3(-t + \tau_4)\right)\right)\right) \quad (31)$$

$$H_5(-t + \tau_4)$$

$$\frac{\partial u_1}{\partial \tau_4} = -H_\infty(-t + \tau_1)\delta_{0.1}\left(-t + \tau_1 H_1\left(-t + \tau_2 H_2\left(-t + \tau_3 H_5(-t + \tau_4)\right)\right)\right) \quad (32)$$

$$\tau_3\delta_5(-t + \tau_4)$$

In this paper, we define the derivative of equations (19), (20), (21), and (22) with respect to $\tau_i$ (for i=1,2,3,4) as representing the "picking" or "recalling" of each event.

If the order of the delta sequence becomes infinite, only equations (23), (24), (26), (27), (29), (30) ,and (31) remain valid. However, if the order of the delta sequence decreases, the other terms come into play. This suggests that recalling a past event can activate other related past events.

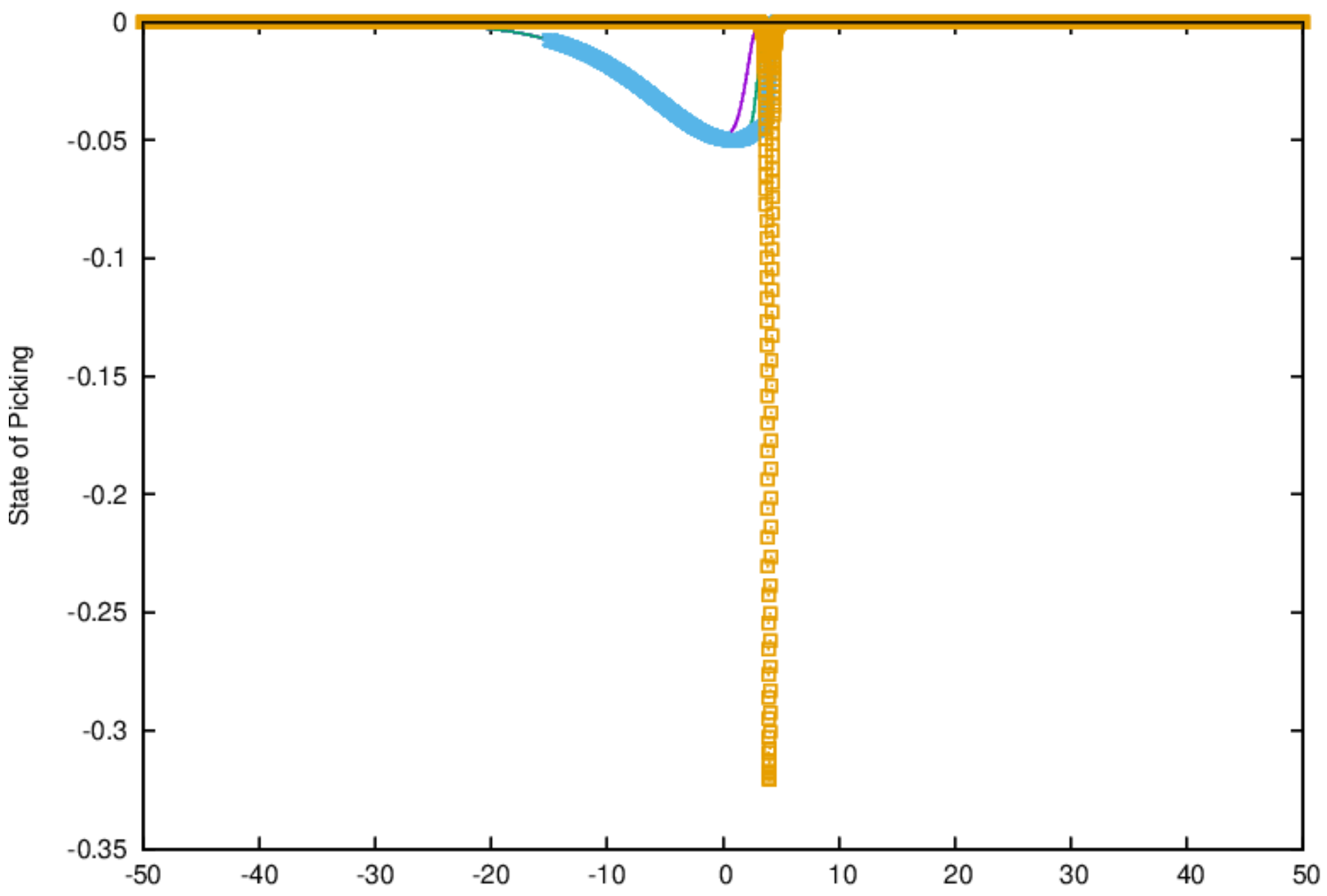

Fig.(7) four different time series of four derivatives from equation(29) to equations(32): the orange corresponds to derivative time series with respect to $\tau_1$ , the blue to $\tau_2$, the purple line to $\tau_3$,and the green to $\tau_4$.

From Figure(7), we note that the orange line corresponds to equation(32) that is distinguished because three terms of equation(32) are reacting together.

Memory fading can be interpreted as a result of the Heaviside and delta sequences having very low order in terms of the derivatives with respect to t and $\tau$ ( Forgetting could be similarly explained).

## Conclusion

In this paper, we have introduced the recursive Heaviside step sequence function as a novel mathematical model for representing human mental states, including thought processes, memory recall, and forgetfulness. By extending the traditional Heaviside step function into a recursive sequence, we have developed a framework that captures the dynamic and layered nature of cognition. This model allows for the differentiation of mental states with respect to both time and event occurrence, providing insight into how thoughts evolve and how memories fade over time.

The recursive Heaviside step sequence function's ability to approximate solutions to the multidimensional inviscid advection equation highlights its versatility in modeling mental processes. When N approaches infinity, the function fully satisfies the advection equation, representing a stable and clearly defined mental state. However, as N decreases, the function reflects more uncertain or forgotten states, demonstrating how memory recall and forgetting can be mathematically modeled.

The selection of N is highly context-dependent, influenced by both the individual's environment and internal cognitive processes. This variability reflects the unique nature of human thought, where certain memories remain vivid while others fade into obscurity. Furthermore, the recursive structure of the Heaviside step sequence allows for the association of ideas, where the recall of one memory can trigger the recollection of related events.

While the current model provides a powerful tool for understanding mental states, there are limitations in its practical implementation. Modern technology has yet to develop the necessary storage and processing capabilities to fully simulate the recursive Heaviside step sequence function in real time. However, this framework offers a promising direction for future research, particularly in fields that intersect applied mathematics, cognitive science, and artificial intelligence.

In conclusion, the recursive Heaviside step sequence function and its associated delta sequence provide a novel and flexible approach to modeling human cognition. By capturing the continuous evolution of mental states, this model offers new insights into how we think, remember, and forget, opening the door for future advancements in both theoretical and applied cognitive studies.

---